\documentclass{article}
\usepackage{spconf,amsmath,graphicx}
\usepackage{algorithm,epsfig,cite,stfloats,bm,amssymb,amsmath,color,subfigure,
graphics,extpfeil,amsthm}
\newtheorem{pro}{Proposition}

\title{How to interconnect for Massive MIMO Self-Calibration?}

\name{Fuqian Yang$^1$, Hanyu Zhu$^1$, Cong Shen$^2$, Linglong Dai$^3$, and Xiliang Luo$^1$}
\address{
$^1$ShanghaiTech University, $^2$University of Science and Technology of China,
$^3$Tsinghua University}
\begin{document}
\ninept
\maketitle
\begin{abstract}
In time-division duplexing (TDD) systems, massive multiple-input multiple-output
(MIMO) relies on the channel reciprocity to obtain the downlink (DL) channel
state information (CSI) with the acquired uplink (UL) CSI at the base station (BS).
However, the mismatches in the radio frequency (RF) analog circuits at
different antennas at the BS break the end-to-end UL and DL channel
reciprocity. To restore the channel reciprocity, it is necessary to calibrate
all the antennas at the BS. This paper addresses the interconnection strategy for
the internal self-calibration at the BS where different antennas are interconnected
via hardware transmission lines. Specifically, the paper reveals the optimality of
the star interconnection and  the daisy chain interconnection
respectively. From the results, we see the star interconnection is the optimal
interconnection strategy when the BS are given the same number of measurements.
On the other hand, the daisy chain interconnection outperforms the star interconnection
when the same amount of time resources are consumed.
Numerical results corroborate our theoretical analyses.
\end{abstract}
\vspace{-0.2cm}
\section{Introduction}\label{SecIntro}
\vspace{-0.1cm}
In massive multiple-input multiple-output (MIMO), a large number of antennas
are installed at the base station (BS) to enhance the system spectral efficiency
\cite{larsson14commag}. To avoid the need to feed back a large amount of downlink
(DL) channel state information (CSI) to the BS as in frequency-division
duplexing (FDD) systems, time-division duplexing (TDD) is typically assumed for
massive MIMO where the channel reciprocity can be exploited to infer the DL CSI
with the uplink (UL) CSI acquired at the BS \cite{Smith2004}.
However, in practice, the transmit and receive
branches are composed of totally different analog circuits. Accordingly, the
radio-frequency (RF) gain of the transmit chain is different from that of the
receive chain at the baseband \cite{LuoMaMIMO}.
These RF gain mismatches destroy the end-to-end TDD channel reciprocity and lead
to severe performance degradation in massive MIMO systems \cite{LuoMaMIMO,wei2016twc,ZhangHardMismatch}. Careful calibration is thus
necessitated to compensate those RF gain mismatches at the RF front ends (FEs)
to restore the end-to-end UL/DL channel reciprocity.

There are two main categories of calibration schemes to compensate the RF gain
mismatches.
One is the ``relative calibration'' and the other one is the ``full calibration''.
The relative calibration was proposed to only restore the end-to-end UL and DL channel
reciprocity without addressing the absolute phase or amplitude coherence
\cite{ShepardArgos}. On the other hand, the full calibration provides full absolute
phase and amplitude coherence between transmitters and receivers
\cite{BenzinInter2017}.

To accomplish either the relative calibration or the full calibration,
either the ``Self-Calibration'' scheme
\cite{Nishimori2001,Liu2006,ShepardArgos,wei2016twc,Vieira2017Proposal,BenzinInter2017}
or the ``Over-The-Air (OTA)'' calibration scheme
\cite{Kalten10futnet,Shi2011,Rogalin2014,LuoCS} can be applied.
By utilizing hardware interconnections with transmission lines
\cite{Nishimori2001,Liu2006,BenzinInter2017} or exploiting
the mutual coupling effects \cite{ShepardArgos,wei2016twc,Vieira2017Proposal},
the self-calibration scheme can be performed by the BS only without invoking helps
from the served mobile stations (MSs) or other antenna arrays.
The OTA calibration scheme is carried out with the help of the assisting MSs or
other antenna arrays \cite{Rogalin2014}. In massive MIMO, the OTA scheme
usually requires a significant amount of CSI feedback from the MSs \cite{LuoCS}.

In this paper, we focus on the internal self-calibration scheme and
seek for the optimal interconnection strategy to wire the antennas
at the BS together with transmission lines. In particular, we analyze the optimality
of the star interconnection and the daisy chain interconnection respectively under
different resource constraints. The derived results in this paper can serve as the
design guidelines for massive MIMO systems.

The rest of the paper is organized as follows.
Section \ref{SecSysModel} gives the system model and performance characterization for
self-calibration.
Section \ref{SecOptStar} analyzes the optimality of the star interconnection strategy
without considering the constraint on the time resources.
Section \ref{SecOptDaisy} shows the optimality of the daisy chain interconnection
strategy when the time resources are limited.
Numerical results are provided in Section \ref{Secnum} and
Section \ref{SecConc} concludes the paper.

\noindent{\it Notations}:
${\sf{Diag}}\{\cdot\}$ denotes the diagonal matrix with the diagonal
elements defined inside the curly brackets.
Notations
${\sf Tr}\{\cdot\}$, $(\cdot)^{T}$, $(\cdot)^H$, $(\cdot)^*$, and $\vert\mathcal{C}\vert$ stand for
matrix trace, transpose operation, Hermitian operation, conjugate operation, and the cardinality of the set $\mathcal{C}$, respectively.
$\mathcal{A}\setminus \mathcal{B}$ means the relative complement of the set $\mathcal{B}$ in the set $\mathcal{A}$. ${\sf mod}(a,b)$ represents modulo operation that finds the remainder after division of $a$ by $b$.
$[\bm A]_{p,q}$ denotes the $(p,q)$-th entry of matrix $\bm A$.

\vspace{-0.2cm}
\section{System Model \& Calibration Performance}\label{SecSysModel}
\vspace{-0.1cm}
\subsection{System Model}
Consider a TDD  multi-user (MU) massive MIMO system with an $M$-antenna BS and $U$ single-antenna MSs.
As many works have shown, only calibration of the BS front-ends is required and
the effects of RF gains at the MSs can be neglected \cite{wei2016twc,BenzinInter2017}.
To obtain calibration coefficients at the BS,
the self-calibration method with hardware circuit connection is considered in this paper \cite{BenzinInter2017}.
The complex-valued transmit and receive RF gains of the antennas at the BS are denoted as ${\{\alpha_m, \beta_m\}}_{m=1}^M$ .
During the calibration phase, the BS antennas transmit
sounding signals over the transmission lines to obtain calibration measurements.
Let $y_{p,q}$ denote the received signal at the $p$-th antenna due to the
transmission from the $q$-th antenna.
Without loss of generality,
the transmitted sounding signal is assumed to be $1$.
It then follows that\vspace{-0.2cm}
\begin{equation}\label{scalarrece}
  y_{p,q}= \beta_p h_{p,q} \alpha_q+n_{p,q},\vspace{-0.2cm}
\end{equation}
where $h_{p,q}$ represents the gain of the calibration channel between the $p$-th antenna and the $q$-th antenna
and $n_{p,q}$ is the additive white Gaussian noise (AWGN)
with zero mean and variance $\sigma_n^2$.
Note that $h_{p,q}=0$ if there is no interconnection wiring between the $p$-th antenna and the $q$-th antenna
and $h_{p,q}=h_{q,p}$ due to the reciprocity of the calibration channel.
Note that every transmission line is bidirectional and two measurements are obtained with each
transmission line.
By stacking all the calibration measurements in
(\ref{scalarrece}) together,
the received signals in matrix-form  is
\vspace{-0.2cm}
\begin{equation}\label{matrixrece}
\bm{Y}=\bm R \bm H \bm T+\bm{N},\vspace{-0.2cm}
\end{equation}
where $[\bm Y]_{p,q}:=y_{p,q}$,
${\bm R}:={\sf Diag}\{\beta_1,\beta_2,\cdots,\beta_M\}$,
${\bm T}:={\sf Diag}\{\alpha_1,\alpha_2,\cdots,\alpha_M\}$,
$[\bm H]_{p,q}:=h_{p,q}$, and $[\bm N]_{p,q}:=n_{p,q}$.

\subsection{Performance Characterization}
\label{SecProCRLB}
In this study, we only focus on the full calibration schemes, but similar results
can also be derived when the relative calibration is considered.
To restore the end-to-end channel reciprocity, the BS only need to
know the values of those transmit and receive RF gains subject to a
common scaling, e.g. $\{s_{\alpha}\alpha_m\}_{m=1}^M$ and
$\{s_{\beta}\beta_m\}_{m=1}^M$. In order to proceed with
our quantitative analyses, we assume there is a ``reference antenna'',
e.g. the $f$-th antenna, whose RF gains: $\alpha_f$ and $\beta_f$ are known \cite{Vieira2017Proposal}.
The other antennas are called ``ordinary antennas'' accordingly.
For a particular interconnection strategy, given all the measurements
$\bm Y$ in (\ref{matrixrece}), the corresponding Cramer-Rao low bounds (CRLBs)
for those unknown calibration coefficients,
i.e. ${\left\{\alpha_m,\beta_m\right\}}_{m=1}^M\setminus
\{\alpha_{f},\beta_{f}\}$, can be derived.
Note these CRLBs serve as
lower bounds for the variances of the estimation errors of all possible
unbiased estimators \cite{Kay1993}.
Let the matrix $\mathcal{A}$ represent the interconnection strategy. Specifically, it is defined as
\vspace{-0.2cm}
\begin{equation}
\mathcal{A}_{p,q} :=
\left\{\begin{array}{cc}
1, &\text{Antenna-}p,q\text{ are interconnected}\\
0, &\text{otherwise}
\end{array} \right..
\end{equation}
And let $\bar{\mathcal{A}}$ be the submatrix obtained by removing the
$f$-th row and the $f$-th column from the interconnection matrix $\mathcal{A}$.
In this paper, we endeavor to find the optimal interconnection strategy or
wiring at the BS which connects different antennas in the most efficient
way to enable the best estimates of the calibration coefficients.
To proceed with our derivations, we make the following assumption:
\begin{itemize}
  \item{\sf AS-1}: All the transmission lines have the same length and damping,
  i.e. $h_{p,q}=h$ when the $p$-th antenna and the $q$-th antenna are
  interconnected.
\end{itemize}
Let $\bm{\theta}$ be a $2(M-1)$-by-$1$ unknown vector defined as
$\bm{\theta}:=[\bm \alpha^T, \bm \beta^T]^T$,
where
$\bm \alpha:=[\alpha_1,...,\alpha_{f-1},\alpha_{f+1},...,\alpha_M]^T$
and
$\bm \beta:=[\beta_1,...,\beta_{f-1}, \beta_{f+1},\ldots, \beta_M]^T$.
According to the signal model
in (\ref{matrixrece}), under AS-1, we have $\bm H = h{\mathcal{A}}$ and
the CRLB matrix for $\bm \theta$ with an interconnection strategy $\mathcal{A}$
can be derived as \cite{Kay1993}
\begin{equation}\label{CRLB_full}
{\sf CRLB}(\bm{\theta}|\mathcal{A})=
  \bm J^{-1}(\bm{\theta}),
\end{equation}
where the Fisher information matrix $\bm J(\bm{\theta})$ is given by
\begin{equation}
\label{FishCtheta2}
\bm J(\bm{\theta})=
\frac{\vert h\vert^{2}}{\sigma_n^2}\cdot
\left[\begin{array}{cc}
\vspace{-0.1cm}   \bm A& \bm D^H\\
   \bm D&\bm B
\end{array} \right],\vspace{-0.3cm}
\end{equation}
with
\begin{equation}
\begin{split}
\bm D &\!:=\!{\sf Diag}\left\{\bm{\beta}\right\}\cdot\bar{\mathcal{A}}\cdot
{\sf Diag}\{\bm{\alpha^*}\},\\
\bm A &\!:=\!{\sf Diag}\Big\{\sum\limits_{i\in\mathcal{C}_1}\!\!|\beta_i|^2, \ldots,\!\!\!\!\!\! \sum\limits_{i\in\mathcal{C}_m, m\ne f}\!\!\!|\beta_i|^2, \ldots,
\sum\limits_{i\in\mathcal{C}_M}\!\!\!|\beta_i|^2\Big\},\\
\bm B &\!:=\!{\sf Diag}\Big\{\sum\limits_{i\in\mathcal{C}_1}\!\!|\alpha_i|^2, \ldots,\!\!\!\!\!\!
\sum\limits_{i\in\mathcal{C}_m, m\ne f}\!\!\!|\alpha_i|^2, \ldots,
\sum\limits_{i\in\mathcal{C}_M}\!\!\!|\alpha_i|^2\Big\},
\end{split}
\end{equation}
and $\mathcal{C}_m$ denotes the set of the indices of the antennas
that are interconnected to the $m$-th antenna directly in this particular
interconnection strategy $\mathcal{A}$.
\begin{figure}[t]
\centering
\epsfig{file=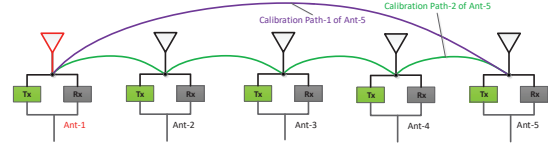,width=0.4\textwidth}
\caption{The interconnection strategy with $5$ antennas.
Antenna-$1$ is chosen as the reference antenna and there are two calibration paths for antenna-$5$.}
\label{fig:pathexample}
\end{figure}
In this paper, we call an interconnection path between one ordinary antenna and the
reference antenna a ``calibration path''. For example, the purple path shown
in Fig. \ref{fig:pathexample} is one calibration path of antenna-$5$.
Note that to be able to estimate all the calibration
coefficients, the chosen interconnection strategy $\mathcal{A}$ must be
``effective'' in the sense that there must be at least one calibration path between each
ordinary antenna and the reference antenna.
Besides, to ensure an effective interconnection strategy,
the BS must be equipped with at least $(M-1)$ transmission lines and at least $2(M-1)$
calibration measurements need to be obtained.
In the following sections, the optimality of different interconnection strategies is analyzed based on the corresponding CRLBs for the unknown calibration coefficients.

\section{Optimality of the Star Interconnection}
\label{SecOptStar}
To gain more insights from the CRLB results in
(\ref{CRLB_full}), we further make the following assumption:
\begin{itemize}
  \item{\sf AS-2}: The transmit and receive RF gains exhibit equal amplitudes, i.e.
  $\vert\alpha_m\vert=a, \vert\beta_m\vert=b$, $\forall m\in[1,M]$.
\end{itemize}
AS-2 is made mainly due to the following concern.
Constant transmit and receive amplitudes ensure identical receive signal-to-noise ratio (SNR)
in the calibration measurements at each BS antenna. The current study only focuses on the
impact of the internal interconnection strategy.
Assuming that the BS has a total budget of $(M-1)$ transmission lines to interconnect different
antenna ports at the BS, under AS-2, closed-form expressions for the CRLBs in (\ref{CRLB_full})
can be derived. Further, the optimal interconnection strategies for internal full
calibration  can be characterized according to the derived analytical results.

\begin{figure}[t]
\centering
\epsfig{file=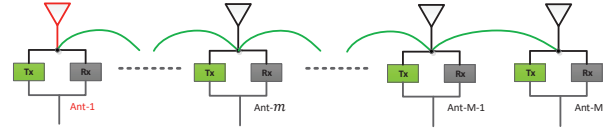,width=0.45\textwidth}
\caption{The daisy chain interconnection strategy with $M$ antennas. ``Ant-f'' is the reference antenna, and number of antennas along the calibration path of the $m$-th antenna in addition to
the reference antenna is $d_m$.}
\label{fig:CalibrationPathEffect}
\end{figure}
\begin{figure}[t]
\centering
\epsfig{file=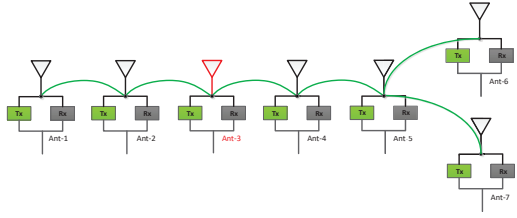,width=0.38\textwidth}
\caption{An interconnection network with $7$ antennas. The antenna-$3$ is chosen as the reference antenna.}
\label{fig:InterconnectionNetwork}
\end{figure}
\begin{figure}[t]
\centering
\epsfig{file=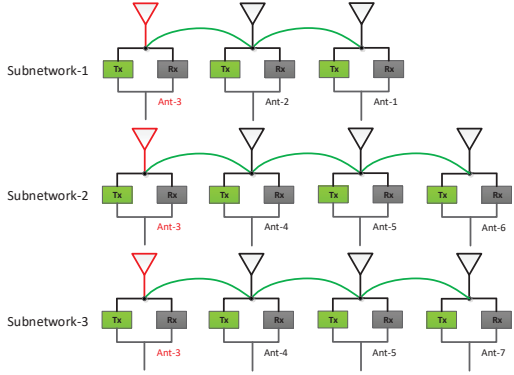,width=0.38\textwidth}
\caption{The three decoupled interconnection subnetworks of the interconnection network in Fig \ref{fig:InterconnectionNetwork}.
The three subnetworks are all the daisy chain interconnection networks.}
\label{fig:DecoupledNetwork}
\end{figure}

To derive the closed-form CRLB expressions, we consider the daisy chain interconnection strategy as shown in Fig. \ref{fig:CalibrationPathEffect}, where the reference is set to $f=1$. According to (\ref{FishCtheta2}), under AS-1 and AS-2, the Fisher information matrix for the daisy chain interconnection strategy with $M$ antennas is given as
\begin{equation}\vspace{-0.2cm}
  \bm J_{\text{daisy}}(\bm \theta)=\frac{\vert h\vert^{2}}{\sigma_n^2}\cdot\left[\begin{array}{cc}
\vspace{-0.1cm}   \bm A& \bm D^H\\
   \bm D&\bm B
\end{array} \right],\vspace{-0.3cm}
\end{equation}
where
\begin{equation}
\begin{split}
  &\bm A=2b^2\cdot {\sf Diag}\{1, 1,\ldots, 1,1/2\},\\
  &\bm B=2a^2\cdot {\sf Diag}\{1, 1,\ldots, 1,1/2\},\\
\end{split}
\end{equation}
\begin{equation}
\begin{split}
  &[\bm D]_{p,q}=\left\{
  \begin{array}{ll}
    \beta_{p+1}\alpha^*_{q+1}, & \vert p-q\vert=1\\
    0, & \text{otherwise}
  \end{array}
   \right.,\\
\end{split}
\end{equation}
with $p,q\in [1,M]$.
Hence, the diagonal elements of  the matrix $\bm J_{\text{daisy}}^{-1}(\bm \theta)$ can be directly obtained according to the inverse of the partitioned matrix $\bm J_{\text{daisy}}(\bm \theta)$. Specifically, we can obtain
\begin{equation}\label{CRLBDaisy0}
\begin{split}
  &[\bm J_{\text{daisy}}^{-1}(\bm \theta)]_{m,m}\!\!=\left\{
  \begin{array}{ll}
  \!m\rho_b,& \!\!\!\!m\in [1, M-1]\\
  \!(m-M+1)\rho_a ,\! \!\!\!\!&m\in [M, 2M-2]
  \end{array}
  \right..
\end{split}
\end{equation}
where $\rho_b: =\sigma_n^2/(b^2\vert h\vert^2)$ and $\rho_a:=\sigma_n^2/(a^2\vert h\vert^2)$.

For an arbitrary effective interconnection strategy with $(M-1)$ transmission lines, the interconnection network can be decoupled into a set of daisy chain interconnection subnetworks.
For example, the interconnection network in Fig. \ref{fig:InterconnectionNetwork} can be decoupled into three daisy chain interconnection subnetworks as shown in Fig. \ref{fig:DecoupledNetwork}.
Since $\alpha_f$, $\beta_f$, and $h$ are assumed to be known, the CRLBs for $\alpha_m$ and
$\beta_m$ are determined by the calibration path of the $m$-th antenna and the SNR in the
corresponding measurements.
Hence, the CRLBs for the calibration
coefficients of an arbitrary interconnection strategy can be obtained by computing the CRLBs
with the decoupled daisy chain interconnection subnetworks independently.
Note the CRLBs with each subnetwork can be directly obtained with the results in
(\ref{CRLBDaisy0}).
We call the number of antennas along the calibration path of an ordinary antenna excluding
the reference antenna as the ``calibration distance''.
Let $d_m$ denote the calibration distance of the $m$-th antenna. According to the results in
(\ref{CRLBDaisy0}), under AS-1 and AS-2, for an arbitrary effective interconnection strategy, the CRLBs for $\alpha_m$ and $\beta_m$, $\forall m\ne f$, can be derived as
\begin{equation}\label{CRLBDaisy}
\begin{split}
  {\sf CRLB}(\alpha_m)&\!=\!d_m\rho_b,
  {\sf CRLB}(\beta_m)\!=\!d_m\rho_a.
\end{split}
\end{equation}
Meanwhile, for an arbitrary effective interconnection strategy, according to the results in (\ref{CRLBDaisy}),
the average CRLBs for $\bm \alpha$ and $\bm \beta$ can be obtained as follows:
\vspace{-0.3cm}
\begin{equation}\label{CRLBAverage}
 \begin{split}
   {\sf CRLB}(\alpha)_{\text{Average}}&=\bar{d}\rho_b,
   {\sf CRLB}(\beta)_{\text{Average}}=\bar{d}\rho_a,
 \end{split}
\end{equation}
where $\bar{d}:=\sum_{m=1}^{M-1}d_m/(M-1)$ represents the average calibration distance.
In particular, for the star interconnection strategy, i.e. all the ordinary antennas are directly interconnected to the reference antenna, the average calibration distance $\bar{d}=1$.
This shows that the star interconnection achieves the smallest average CRLB.
In summary, we can establish the following result.
\begin{pro}\label{pro1}
Considering a BS with $M$ antennas interconnected with $(M-1)$ transmission lines, assuming only total $2(M-1)$ measurements are available, under AS-1 and AS-2, the star interconnection minimizes the average CRLB for all the unknown calibration coefficients during internal self-calibration.
\end{pro}
Proposition \ref{pro1} indicates the star interconnection strategy is the optimal
interconnection when the BS has $2(M-1)$ measurements.
In next section, we will analyze the optimality of the daisy chain interconnection
strategy under limited time resources.

\section{Optimality of the Daisy Chain}
\label{SecOptDaisy}

As mentioned in \cite{BenzinInter2017}, the daisy chain interconnection
strategy requires fewer time resources to collect the $2(M-1)$ measurements.
However, the authors in \cite{BenzinInter2017} only demonstrated that the daisy
chain interconnection could outperform the star interconnection with numerical
simulations. In this section, we will prove the optimality of the daisy chain
interconnection and the corresponding condition.

When $M=2$, there is only one unique interconnection strategy.
In the following analyses, it is thus assumed that $M\ge 3$.
To characterize the constraint on the time resources,
we assume that each measurement of the sounding signal consumes $T$ seconds.
To obtain $2(M-1)$ measurements, with the star interconnection strategy, we need
$T_{\text{star}}=2(M-1)T$ seconds.
However, with the daisy chain interconnection strategy, we only need
$T_{\text{daisy}}=4T$ seconds to collect the same amount of measurements due
to the fact that these measurements can be performed in parallel
\cite{BenzinInter2017}. Consider an arbitrary effective interconnection strategy
with $(M-1)$ transmission lines. Let $N_m$ be the number of antennas that directly
interconnected to the $m$-th antenna. Define $N_{\max}:=\max\{N_m|m=1,...,M\}$.
Note that $2\le N_{\max}\le M-1$ when $M\ge 3$.
To obtain $2(M-1)$ measurements, we need at least $2N_{\max}T$ seconds.
It can be observed that $N_{\max}=2$ if and only if the interconnection is the
daisy chain interconnection. Further, we see $N_{\max}=M-1$ if and only if the
interconnection is the star interconnection. We can first have the following
proposition summarizing our findings.

\begin{pro}\label{pro2}
For a BS with $M\ge 3$ antennas interconnected with $(M-1)$ transmission lines,
we need $T_{\text{arb}}$ seconds to obtain $2(M-1)$ calibration measurements
with an arbitrary interconnection strategy.
The required time $T_{\text{arb}}$ satisfies the following condition:
\begin{equation}
	4T\le T_{\text{arb}}\le 2(M-1)T,
\end{equation}
where the first equality holds if and only if the interconnection is the daisy chain interconnection, and the second equality holds if and only if the interconnection is the star interconnection.
\end{pro}

Assume that the BS has a total budget of $T_{\text{star}}=2(M-1)T$ seconds to collect
the calibration measurements. For an arbitrary effective interconnection strategy,
we can utilize the additional $T_{d}:=(T_{\text{star}}-T_{\text{arb}})$ seconds
to acquire additional measurements.
Define $I:=\lfloor \frac{T_{\text{star}}}{T_{\text{arb}}}\rfloor\ge 1$
and $F:={\sf mod}(T_{\text{star}}, T_{\text{arb}})$.
In the following derivations, it is assumed that $(I-1)T_{\text{arb}}$ seconds are used to obtain
additional $(I-1)$ independent $2(M-1)$ measurements for all the unknown calibration
coefficients and the remaining $F$ seconds are not considered\footnote{The remaining $F$ seconds can further be used to improve the total calibration performance, and the effects of the remaining $F$ seconds can also be analyzed with similar methods in the paper.}.
As a result, the average CRLBs for $\bm \alpha$ and $\bm \beta$ with an arbitrary effective interconnection strategy are give by
\begin{equation}
  \begin{split}
    {\sf CRLB}(\alpha)_{\text{Average}}&=\rho_b\bar{d}/I,  {\sf CRLB}(\beta)_{\text{Average}}=\rho_a\bar{d}/I.
  \end{split}
\end{equation}
According to Proposition \ref{pro2}, the daisy chain interconnection strategy consumes
the least amount of time resources to collect the measurements and the additional time
resources can be utilized to improve the calibration performance. For the daisy chain
interconnection strategy, as defined in (\ref{CRLBAverage}), the average calibration
distance $\bar{d}$ is given by
\vspace{-0.1cm}
\begin{equation}\label{dmeanDaisy}
    \bar{d}= \frac{(M-2f)}{2}+\frac{(f-1)^2}{M-1}+1.
    \vspace{-0.1cm}
\end{equation}
Meanwhile, we have $I=\lfloor\frac{(M-1)}{2}\rfloor$ when $2(M-1)T$ seconds are available
to acquire the $2(M-1)$ calibration measurements.
Note that the average calibration distance $\bar{d}$ in (\ref{dmeanDaisy}) is minimized
when $f=\frac{M+1}{2}$. Accordingly, the average CRLB is minimized when
$f=\lfloor\frac{M+1}{2}\rfloor$ for the daisy chain interconnection strategy.
Now we can establish the following results.

\begin{pro}\label{pro3}
For a BS with $M$ antennas interconnected with $(M-1)$ transmission lines,
assuming total $2(M-1)T$ seconds are available to acquire the $2(M-1)$ calibration measurements,
under AS-1 and AS-2, the daisy chain interconnection strategy gives the best calibration
performance when $f=\lfloor\frac{M+1}{2}\rfloor$. The corresponding average
CRLBs are
\begin{equation}
\begin{split}
{\sf CRLB}(\alpha)_{\text{Average}}&=\rho_b\bar{d}/I,
{\sf CRLB}(\beta)_{\text{Average}}=\rho_a\bar{d}/I,
\end{split}
\end{equation}
where $\bar{d}/I=(M+1)/(2M-2)$ when $M$ is odd
and $\bar{d}/I=(M^2)/(2M^2-6M+4)$ when $M$ is even.
\end{pro}

From Proposition \ref{pro3}, we see $\bar{d}/I< 1$ when $M\ge 5$.
In other words, the daisy chain interconnection outperforms the star interconnection
when $M\ge 5$. Further, it can be seen that the value of $\bar{d}/I$ becomes
smaller as the number of antennas increases when $M\ge 5$.
As $M\rightarrow \infty$, for the daisy chain interconnection strategy, the
asymptotic average CRLBs for $\bm \alpha$ and $\bm\beta$ are given by
\begin{equation}\label{CRLBAsy}
\lim_{M\rightarrow\infty}\!\!{\sf CRLB}(\alpha)_{\text{Average}}\!=\rho_b/2,
\lim_{M\rightarrow\infty}\!\!{\sf CRLB}(\beta)_{\text{Average}}\!=\rho_a/2.
\end{equation}
These asymptotic average CRLBs are just half of the corresponding results for the star
interconnection strategy.
The above results show that,
with a total budget of $2(M-1)T$ seconds for calibration, the daisy chain interconnection
strategy outperforms the star interconnection when $M\ge 5$. The relative performance gain
decreases as $M$ goes large and the average CRLBs are bounded by (\ref{CRLBAsy}).

\vspace{-0.2cm}
\section{Numerical Results}\label{Secnum}
\vspace{-0.1cm}

\begin{figure}[t]
\centering
\epsfig{file=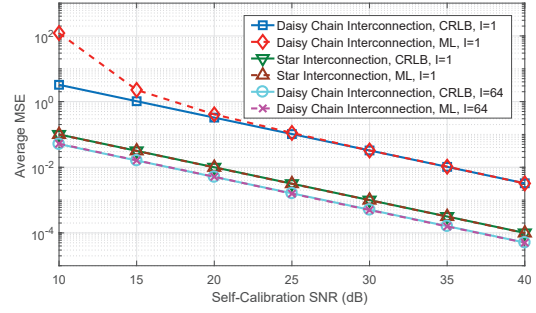,width=0.45\textwidth}
\caption{Full calibration for different interconnection strategies.
(``Star Interconnection'': star interconnection strategy is used for full
calibration at the BS;
``Daisy Chain Interconnection'': daisy chain interconnection is used for
full calibration at the BS;
``CRLB'': average CRLB over all the unknown calibration coefficients;
``ML'': simulated average MSE of all the estimated calibration coefficients
for different interconnection strategies with the maximum-likelihood
(ML) estimator;
Symbol $I$ in the legend means $2I(M-1)$ independent measurements are
obtained.)}
\label{fig:MSE}
\vspace{-0.3cm}
\end{figure}

In this section, numerical results are provided to verify our analytical results.
In our simulations, we compare the star interconnection and the daisy chain
interconnection for self-calibration at the BS.
In order to align with our analytical CRLBs in (\ref{CRLBDaisy}), we assume
that the RF gains of the reference antenna , i.e. $\alpha_f$ and $\beta_f$,
are given and fixed. In the meantime, the interconnection channel $h$ is
also assumed to be known. Note the interconnection channel $h$ is
time-invariant and can be estimated in advance. Then we can obtain the ML estimates of
the calibration coefficients, i.e. $\alpha_m$ and $\beta_m$, for any effective
interconnection strategy implemented at the BS.
Some key parameters assumed in the simulations are listed as follows.
\begin{itemize}
\item The number of antennas at the BS is set to $M=129$;
\item The amplitudes of the transmit and receive RF gains are equal to $1$,
i.e. $|\alpha_m|=|\beta_m|=1,\forall m\in[1,M]$.
The phases of the RF gains are uniformly distributed within $[-\pi, \pi]$;
\item The transmitted sounding signal is equal to $1$;
\item The reference antenna is the $64$-th antenna, i.e. $f=64$;
\item The SNR in the calibration measurements varies from $10$dB to $40$dB.
\end{itemize}
In Fig. \ref{fig:MSE}, the average CRLB and the simulated average mean-square-error
(MSE) of all unknown calibration coefficients are compared under different constraints.
The results show that the star interconnection outperforms the other interconnection
strategies when $I=1$, i.e. $2(M-1)$ calibration measurements are available.
However, when we have $2(M-1)T$ seconds time resources, the daisy chain interconnection
strategy shows better calibration performance than the star interconnection strategy.

\vspace{-0.2cm}
\section{Conclusions}\label{SecConc}
\vspace{-0.1cm}

In this paper, we have studied the interconnection strategy for internal
self-calibration in massive MIMO systems. Based on the derived CRLBs for the
unknown calibration coefficients, on the one hand, we have shown that the star
interconnection is the optimal strategy to interconnect the antennas at the BS
for internal self-calibration when the BS has $2(M-1)$ measurements.
On the other hand, when the number of antennas becomes large, the daisy chain
interconnection outperforms the star interconnection when the BS only has
$2(M-1)T$ seconds to collect the calibration measurements.

\bibliographystyle{IEEE}

\begin{thebibliography}{11}
\bibitem{larsson14commag}
E.~G.~Larsson, O.~Edfors, F.~Tufvesson, and T.~L.~Marzetta, ``Massive
MIMO for next generation wireless systems,'' \emph{IEEE Commun. Mag.},
vol.~52, no.~2, pp. 186--195, Feb. 2014.

\bibitem{Smith2004}
G. S. Smith, ``A direct derivation of a single-antenna reciprocity relation
for the time domain,'' \emph{IEEE Trans. Antennas Propag.}, vol. 52, no. 6,
pp. 1568--1577, Jun. 2004.

\bibitem{LuoMaMIMO}
X. Luo, ``Multiuser massive MIMO performance with calibration errors,'' \emph{IEEE Trans.
Wireless Commun.}, vol. 15, no. 7, pp. 4521--4534, Jul. 2016.

\bibitem{ZhangHardMismatch}
W. Zhang, H. Ren, C. Pan, M. Chen, R. C. Lamare, B. Du, and J. Dai, ``Large-scale antenna
systems with UL/DL hardware mismatch: Achievable rates analysis and calibration,''
\emph{IEEE Trans. Commun.}, vol. 63, no. 4, pp. 1216--1229, Apr. 2015.

\bibitem{wei2016twc}
H. Wei, D. Wang, H. Zhu, J. Wang, S. Sun, and X. You, ``Mutual
coupling calibration for multiuser massive MIMO systems,'' \emph{IEEE
Trans. Wireless Commun.}, vol. 15, no. 1, pp. 606--619, Jan. 2016.

\bibitem{ShepardArgos}
C. Shepard, H. Yu, N. Anand, L. E. Li, T. Marzetta, R. Yang, and L. Zhong, ``Argos:
Practical many-antenna base stations,'' in \emph{Proc. ACM Int. Conf. Mobile Comput.
Netw. (Mobicom)}, Istanbul, Turkey, Aug. 2012, pp. 1--12.

\bibitem{BenzinInter2017}
A. Benzin and G. Caire, ``Internal self-calibration methods for large scale array
transceiver software-defined radios,'' in \emph{Proc. IEEE  Int. ITG Workshop Smart
Antennas}, Berlin, Germany, Mar. 2017, pp. 1--8.



\bibitem{Nishimori2001}
K. Nishimori, K. Cho, Y. Takatori, and T. Hori, ``Automatic calibration
method using transmitting signals of an adaptive array for TDD systems,''
 \emph{IEEE Trans. Veh. Technol.}, vol. 50, no. 6, pp. 1636--1640, Nov. 2001.
\newpage
\bibitem{Liu2006}
J. Liu, G. Vandersteen, J. Craninckx, M. Libois, M. Wouters, F. Petr\'e, and A. Barel,
``A novel and low-cost analog frond-end mismatch calibration scheme for MIMO-OFDM WLANs,''
in \emph{Proc. IEEE Radio Wireless Symp.,} San Diego, CA, Oct. 2006, pp. 219--222.

\bibitem{Vieira2017Proposal}
J. Vieira, F. Rusek, O. Edfors, S. Malkowsky, L. Liu, and F. Tufvesson, ``Reciprocity
calibration for massive MIMO: Proposal, Modeling, and Validation,'' \emph{ IEEE Trans.
Wireless Commun.}, vol. 16, no. 5, pp. 3042--3056, May 2017.

\bibitem{Kalten10futnet}
F. Kaltenberger, H. Jiang, M. Guillaud, and R. Knopp, ``Relative channel
reciprocity calibration in MIMO/TDD systems,'' in \emph{Proc. IEEE Future
Netw. Mobile Summit,} Florence, Italy, Jun. 2010, pp. 1--10.

\bibitem{Shi2011}
J. Shi, Q. Luo, and M. You, ``An efficient method for enhancing TDD
over the air reciprocity calibration,'' in \emph{Proc. IEEE Wireless Commun. Networking
Conf.}, Cancun, Quintana Roo, Mar. 2011, pp. 339--344.

\bibitem{Rogalin2014}
R. Rogalin, O. Y. Bursalioglu, H. Papadopoulos, G. Caire, A. F. Molisch,
A. Michaloliakos, V. Balan, and K. Psounis, ``Scalable synchronizaiton
and reciprocity calibration for distributed multiuser MIMO,''  \emph{IEEE
Trans. Wireless Commun.}, vol. 13, no. 4, Apr. 2014.

\bibitem{LuoCS}
X. Luo, ``Robust large scale calibration for massive MIMO,'' in \emph{Proc. IEEE Global
Commun. Conf. (GLOBECOM)}, San Diego, CA, Dec. 2015, pp. 1--6.

\bibitem{Kay1993}
S. M. Kay, \emph{Fundamentals of Statistical Signal Processing: Estimation Theory.} Upper
Saddle River, New Jersey, USA: Prentice Hall, 1993.



\end{thebibliography}

\end{document}